\documentclass{PoS}

\PoS{PoS(HEP2005)354}

\title{The Electroweak Phase Transition in Models with Gauge-Higgs Unification}

\ShortTitle{The EWPT with Gauge-Higgs Unification}

\author{\speaker{Marco Serone}\\
        SISSA and INFN, via Beirut 2, I-34014 Italy\\
        E-mail: \email{serone@sissa.it}}

\author{Giuliano Panico\\
SISSA and INFN, via Beirut 2, I-34014 Italy\\
        E-mail: \email{panico@sissa.it}}

\abstract{The dynamics of five dimensional Wilson line phases at finite
temperature is studied in the one-loop approximation.
We show that at temperatures of order $T\sim 1/L$, where $L$
is the length of the compact space, the gauge symmetry is always restored and
the electroweak phase transition appears to be of first order.

We focus on a specific model where the Wilson line phase is
identified with the Higgs field (gauge-Higgs unification).
The transition is of first order even for values of the Higgs mass
above the current experimental limit.
If large localized gauge kinetic terms are present, the transition
might be strong enough to give baryogenesis at the electroweak transition.}

\FullConference{International Europhysics Conference on High Energy Physics\\
         July 21st - 27th 2005\\
         Lisboa, Portugal}

\def\bea{\begin{eqnarray}} \def\eea{\end{eqnarray}}
\def\be{\begin{equation}} \def\ee{\end{equation}} 
  \def\Z{{\bf Z}}

\begin{document}

\section{Introduction}

Theories with compact extra dimensions at the TeV scale \cite{Antoniadis}
offer the possibility of identifying the Higgs field with the internal component
of a gauge field.\footnote{See \cite{Serone:2005ds} for a brief overview and earlier references.}
The Electroweak Symmetry Breaking (EWSB)
is equivalent to a Wilson line symmetry breaking, since the vacuum expectation value (VEV) of the
Higgs field is proportional to a Wilson line phase along the compact extra dimensions.
Such theories with gauge-Higgs unification (GHU) can provide
a solution to the gauge hierarchy problem. 
In the minimal five-dimensional (5D) case, which seems the most interesting framework,
interesting potentially realistic models of GHU
have recently appeared in \cite{GHU}.

In this work we study if and how in such models an electroweak phase transition (EWPT) arises
at some critical temperature $T_c$, above which the symmetry is restored.
One of the main motivations
to perform this study is related to the possibility of having a successful baryogenesis
if a strong first-order phase transition occurs. 
We focus on models in flat space, 
analyzing first the dynamics of 5D Wilson line phases at finite temperature in general and then
considering in detail the study of the Higgs potential in a specific class
of $S^1/\Z_2$ orbifold models \cite{Scrucca:2003ra}.

At one-loop level, we find that the gauge symmetry broken
by the Wilson lines is restored at temperatures of order
$T\sim 1/L$, where $L$ is the length of the covering compact space \cite{Ho:1990xz}. 
The transition is typically of first-order, due to the presence
of a term, cubic in the Wilson line phase, given by massless 5D bosons \cite{Pan}.
An analysis of the EWPT in the model of \cite{Scrucca:2003ra} shows that 
for low values of the Higgs mass, $M_H\lesssim 20$ GeV,
the transition is strongly of first-order, with a strength that is inversely proportional to the Higgs mass,
similarly to what happens in the SM (see fig.~\ref{figCrTempBFerm1}).
In the model of \cite{Scrucca:2003ra}, realistic values of the Higgs mass can be obtained only by
considering generalizations of the minimal model, introducing 5D bulk fermions
in large representations of the gauge group or large localized gauge kinetic terms.\footnote{Both possibilities do not
actually give rise to a phenomenologically acceptable model.
The nature of the EWPT, however, do not depend much on the various
issues that rule out the above models.} 
In the former case, the first order phase transition becomes very weak, whereas
in the latter the strength of the transition is considerably larger.

\section{The Phase Transition in the SM}

The total tree-level and one-loop SM Higgs potential at high temperatures, for $M_H<M_W$, 
can be written as  \cite{Quiros:1999jp}
\be
V_{tot}(T,H) \simeq  D(T^2-T_0^2) H^2 - E T H^3 + \frac 14 \lambda H^4 \,,
\label{V-eff2}
\ee
where $T_0$, $D$ and $E$ are three constants which encode the contributions 
of the gauge bosons and the top quark and
$\lambda$ is the tree-level Higgs quartic coupling.
The coupling $E$ is induced only by the Matsubara zero mode,
which is present only for bosons, and determines
the nature of the phase transition. If $E=0$, one gets a second-order
phase transition, whereas for $E > 0$ one finds a first-order phase transition with a strength
proportional to $|H(T_c)|/T_c \simeq E/\lambda$.
Since $E/\lambda \sim 1/M_H^2$, the transition is weaker and weaker for increasing values
of $M_H$.
The parameter $|H(T_C)|/T_C$ is the crucial parameter to look at if one wants to get baryogenesis
at the EWPT.  For the SM, the requirement is $|H(T_C)|/T_C> 1$.
It turns out that only for $M_H$ significantly lower than $M_W$ the above one-loop computation can be trusted.
Around the critical temperature, as $M_H$ approaches $M_W$,
perturbation theory is less and less reliable, and
for $M_{H}\gtrsim M_W$, perturbation theory breaks down. 
Lattice computations seem to indicate that for $M_H\gtrsim M_W$ the SM has a crossover, ruling then out
baryogenesis at the SM EWPT.

\section{Wilson Lines at Finite Temperature}

Consider, for simplicity, a single Wilson line phase $\alpha$  and
one massless 5D gauge boson and fermion, with charges $q_B$ and $q_F$ ($q_F>q_B$) with respect to $\alpha$.
The effective potential $V$ at $T=0$  has a minimum at $\alpha_{min} \simeq 1/(2q_F)$, 
where the ``Higgs mass'', neglecting the bosonic contribution, 
is approximately given by\footnote{The precise
coefficient relating $\alpha$ to the Higgs VEV $H$ is model dependent.
For definiteness, we have taken here and in the following
the one appearing in \cite{Scrucca:2003ra}.} $M_H^2 =  (g_4 R/2)^2
V^{\prime\prime}(\alpha_{min})\sim 24
g_4^2 q_F^2/(16\pi^2 L^2)$.
At high temperatures $TL>1$, $V$ can roughly be
written, by expanding in $\alpha$, as \cite{Pan}
\be
\frac{L^4}{\pi^2} V(T,\alpha) \simeq a(x) \,\alpha^2 - b(x) \, \alpha^3 + c(x) \,\alpha^4\,, \ \ \ \ \ 
\left\{\begin{array}{l} a(x) = q_B^2 x - 8 q_F^2 \sqrt{2} x^{5/2}e^{-\pi x}  \\
b(x) = 2 x q_B^3  \\
c(x) = q_B^4 x + \frac83 q_F^4 \sqrt{2}\pi^2 x^{5/2}e^{-\pi x}  
\end{array}
\right.
\label{Pot-gen}
\ee
where $x=LT$.
Eq.(\ref{Pot-gen}) is valid for $0\leq \alpha \leq 1/(2q_F)$, which is the
relevant range in $\alpha$ for the study of the phase transition.
The latter is of first-order and occurs with a critical temperature $T_c\sim 1/L$.
At $T=T_C$, $\alpha_{min}(x_C) = b(x_C)/(2 c(x_C))\simeq 6 q_B/(\pi^2 q_F^2)$.
In terms of $|H(T_C)|/T_C = 2\alpha_{min}(T_C) /(g_4 R T_C)$, we get
$|H(T_C)|/T_C \sim q_B/q_F^2$.
The strength of the first-order phase transition is inversely proportional to $q_F^2$
and hence to the value of the squared Higgs mass.

\section{The Phase Transition in a Model with Gauge-Higgs Unification}

\begin{figure}[h!]
\begin{center}
\begin{tabular}{cc}
\includegraphics[width=.45\textwidth]{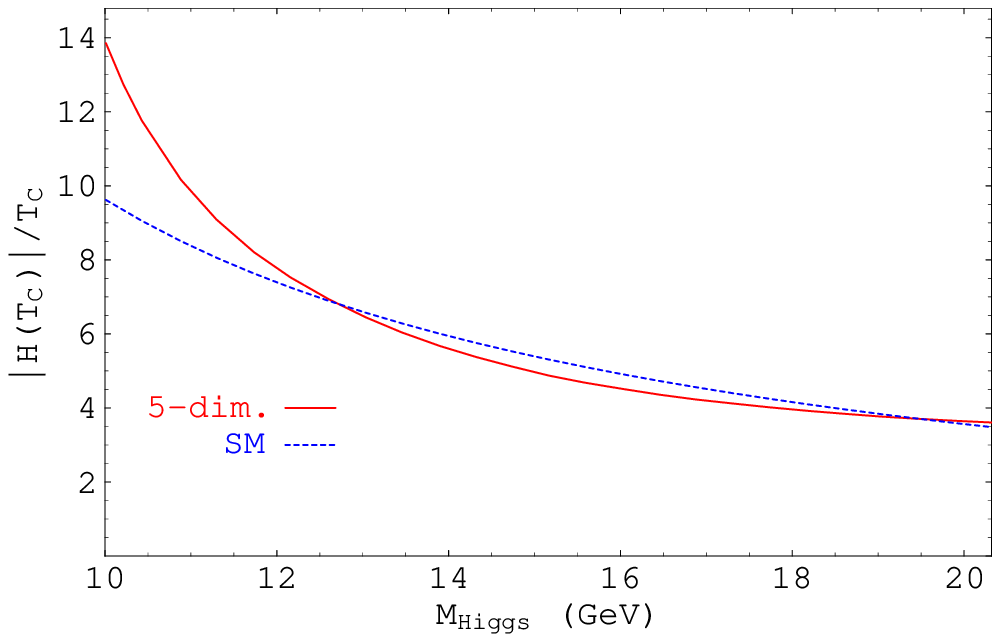}
&\includegraphics[width=.45\textwidth]{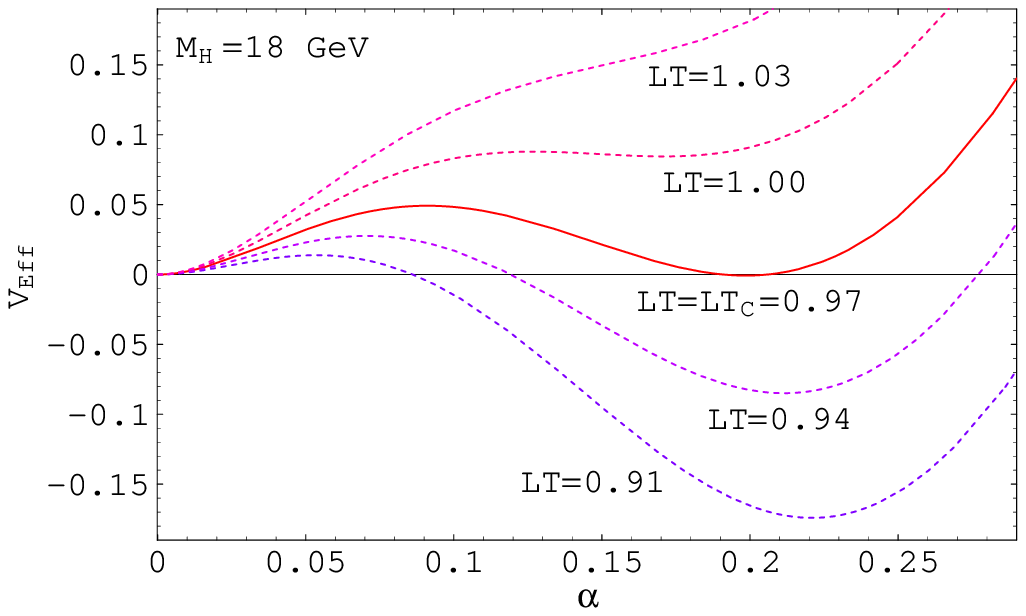}
\end{tabular}
\caption{(Left) Phase transition strength and (right) detail of the potential near
the phase transition (minimal model).}
\label{figCrTempBFerm1}
\end{center}
\end{figure}

We consider here the EWPT in the model of \cite{Scrucca:2003ra}, where we refer for 
a detailed description.
All the results have been obtained by a numerical computation of the 
one-loop Higgs potential (see also \cite{Maru:2005jy}).
In the minimal model of \cite{Scrucca:2003ra}, 
the top mass is too low and it has been fixed to
$M_{top}=45 \,\mathrm{GeV}$.
In agreement with the previous considerations, the EWPT is of first order,
with a critical temperature of order $1/L$ (see fig.~\ref{figCrTempBFerm1} (right)).
For comparison, in fig.~\ref{figCrTempBFerm1} (left)
we plot $|H(T_C)|/T_C$ as a function
of the Higgs mass for both the 5D model and the one-loop SM potential, with $M_{top}=45 \,\mathrm{GeV}$.
The phase transition is strongly first order,
as expected for such low values of $M_H$.

The problem of a too low value for the Higgs mass in the minimal model of \cite{Scrucca:2003ra}
can be solved by the introduction of additional bulk fermions, 
in high rank representations of the underlying $SU(3)_w$ gauge group.
By adding a massive bulk fermion in the symmetric rank $8$ representation of $SU(3)_w$ 
and still fixing the top mass to $M_{top}=45\,\mathrm{GeV}$,
the first-order phase transition becomes much weaker, with $|H(T_C)|/T_C\sim 0.13$ for $110 \leq M_H \leq 150$ GeV.
This can be understood by 
noting that $V$ is now dominated by the high rank fermion and that 
the strength of the phase transition decreases with the rank, as explained in the last section.
No comparison with the SM is given, since for such values of the Higgs mass
perturbation theory breaks down in the SM close to $T_C$.
The introduction of localized gauge kinetic terms represents another way
to get realistic values for the Higgs mass. 
In the notation of \cite{Scrucca:2003ra},
we take $c_1 \equiv c = 6$, $c_2 = 0$ and we fix $M_{top}= 110$ GeV. 
The phase transition is moderately strong of first order, with
$|H(T_C)|/T_C\sim 0.7$ for $110 \leq M_H \leq 170$.
The behaviour of the phase transition for large values of $c$ is analytically hard to be studied.
At the critical temperature, the ratio between $b(x)$
and $c(x)$ (see eq.(\ref{Pot-gen})) depends on the bulk fermion
charges like $1/q_F^2$, but it has a milder dependence on $c$, like $1/\sqrt{c}$.
This explains why in the present case, in which the bulk fermion
charges are small ($q_F^{max}=2$), the first order phase transition is
considerably stronger than in the case with high rank bulk fermions (in which
$q_F^{max}=8$).

All these results are based on one-loop perturbative studies, for any value of $M_{H}< M_W$ or
$M_{H}\geq M_W$.  In light of the breakdown of perturbation theory for
$M_{H}\gtrsim M_W$ in the SM, it is natural to ask if and to what extent one can trust
our results. 
It is possible to give
an estimate of the relevance of the leading higher loop corrections in a
5D Wilson line based toy model \cite{Pan}. 
Due to the good UV properties of the potential of Wilson line phases, 
it has been shown that higher order diagrams give a negligible contribution 
for $T\sim 1/L$.
We believe that these results provide a strong evidence that perturbation theory in 5D models
with GHU is valid around the critical temperature.

\end{document}